\documentclass[pra,twocolumn,showpacs,preprintnumbers,superscriptaddress]{revtex4-2}
\usepackage{xcolor}
\usepackage{times}
\usepackage{bm}
\usepackage{float}
\usepackage{graphicx}
\usepackage{amsbsy}
\usepackage{amsmath}
\usepackage{amsfonts}
\usepackage{amsthm}
\usepackage{placeins}
\usepackage{autobreak}
\usepackage{booktabs}
\usepackage{multirow}
\usepackage[margin=1in]{geometry}

\begin{document}
	
\allowdisplaybreaks
	
\theoremstyle{plain}
\newtheorem{theorem}{Theorem}
\newtheorem{lemma}[theorem]{Lemma}
\newtheorem{corollary}[theorem]{Corollary}
\newtheorem{proposition}[theorem]{Proposition}
\newtheorem{conjecture}[theorem]{Conjecture}
	
\theoremstyle{definition}
\newtheorem{definition}[theorem]{Definition}
	
\theoremstyle{remark}
\newtheorem*{remark}{Remark}
\newtheorem{example}{Example}
\title{Estimation of Power in the Controlled Quantum Teleportation through the Witness Operator}
\author{Anuma Garg, Satyabrata Adhikari}
\email{anumagarg\_phd2k18@dtu.ac.in, satyabrata@dtu.ac.in} \affiliation{Delhi Technological University, Delhi-110042, Delhi, India}
	
\begin{abstract}
\centerline{Abstract}
Controlled quantum teleportation (CQT) can be considered as a variant of quantum teleportation in which three parties are involved where one party acts as the controller. The usability of the CQT scheme depends on two types of fidelities viz. conditioned fidelity and non-conditioned fidelity. The difference between these fidelities may be termed as power of the controller and it plays a vital role in the CQT scheme. Thus, our aim is to estimate the power of the controller in such a way so that its estimated value can be obtained in an experiment. To achieve our goal, we have constructed a witness operator and have shown that its expected value may be used in the estimation of the lower bound of the power of the controller. Furthermore, we have shown that it is possible to make the standard W state useful in the CQT scheme if one of its qubits either passes through the amplitude damping channel or the phase damping channel. We have also shown that the phase damping channel performs better than the amplitude damping channel in the sense of generating more power of the controller in the CQT scheme.      
\end{abstract}
\pacs{03.67.Hk, 03.67.-a} 
\maketitle
	
\section{introduction}
\noindent The process of transferring an
unknown quantum state between two parties at two distant locations without transferring the physical information about the unknown quantum state
itself is known as quantum teleportation \cite{nielsen,wilde,bennett2}. This means that neither any physical information about the state is transferred nor a swap-operation between the sender and the receiver is performed. Teleportation protocol makes use of the non-local correlations generated by using an entangled pair between the sender and the
receiver, and the exchange of classical information between them. This concept plays a
central role in quantum communication using quantum repeaters \cite{gisin1, briegel} and can also be used to implement logic
gates for universal quantum computation \cite{gottesman}. Quantum teleportation was also demonstrated experimentally \cite{bouwmeister,kwiat,michler,boschi}. \\
In standard quantum teleportation prtocol, Alice the sender and Bob the receiver share a maximally two-qubit entangled state. Alice is then supposed to send an arbitrary single qubit state to Bob using the shared maximally entangled state. To execute the quantum teleportation protocol, Alice makes Bell state measurement on the composite system of two qubit with her possession. After the measurement, a state will be projected at Bob's location. Alice communicates her measurement result to Bob by sending the two classical bits $\{00,01,10,11\}$ using a classical channel. According to the received two classical bits, Bob may apply the appropriate unitary operator on his qubit to retrieve the state sent by Alice. The faithfulness of this quantum teleportation scheme is quantified by the fidelity, which may be defined as the maximum overlap between the state to be teleported by Alice and the state at Bob's site. In case of conventional teleportation scheme, the average fidelity is unity. But if either the non-maximally entangled pure two-qubit state or a mixed two-qubit entangled state is used as a resource in the teleportation protocol then the average fidelity of teleportation will always be less than unity, provided the filtering operation is not allowed.\\ 
Quantum teleportation using three qubit state as a resource state was introduced by Karlsson et al. \cite{karlsson}. It is a variant of teleportation in which three members such as Alice (A), Bob (B) and Charlie (C) are participating with one qubit each. Later, this type of quantum teleportation protocol is popularly called as the controlled quantum teleportation (CQT). The CQT scheme proposed by Karlsson et al. \cite{karlsson}, have used maximally entangled three-qubit GHZ state as a shared three-qubit state while on the other hand, Gao et al. \cite{gao} have shown that non-maximally pure three-qubit entangled states such as maximal slice state can be used in CQT with unit probability and unit fidelity. In 2014, Li et al. \cite{li2014} have focussed on the role of controller in the CQT scheme and also defined the power of the controller. Later, they have generalised their CQT protocol for multiqubit pure system \cite{li2015}. K. Jeong et al. \cite{jeong} also have considered the $n-$ qubit GHZ and W state as a channel for CQT and studied it using minimal control power. They have shown that the maximal values of the minimal control power may be obtained for $n-$ qubit GHZ and W class of states. In another work, Artur et al. \cite{Artur} have derived the lower and upper bound of the control power in terms of the three-tangle of any pure three-qubit states. Moreover, Paulson et al. also studied the CQT protocol using the symmetric mixed state such as $X-$ matrix states. Interestingly, it was found that the optimum controlled quantum teleportation fidelity has been obtained for non-maximally entangled mixed $X-$ states \cite{paulson}. The CQT scheme also has been investigated using high-dimensional tripartite standard GHZ and other GHZ classes of states \cite{wang}. Recently, controlled quantum teleportation was  experimentally realized using cluster states \cite{kumar}.    
The potential application of controlled quantum teleportation may be found in quantum computing algorithms, quantum communication protocols, and quantum error correction schemes \cite{rau}. The concept of CQT may also be used in quantum networks \cite{hamdoun}, entanglement swapping \cite{jun}, quantum reapters \cite{sango}, and quantum key distribution \cite{sayan}.\\
We are now in a position to discuss the detailed scheme of CQT. In the CQT scheme, we may consider that Alice, Bob and Charlie shared a three-qubit pure/mixed state described by the density operator $\rho_{ABC}$. We assume throughout the paper that Charlie act as a controller who perform von-Neumann measurement on his qubit. A single qubit von-Neumann measurement in the computational basis may be described as \{$\pi_{k}=|k\rangle\langle k|,k=0,1$\}. In general, a single qubit measurement operator in an arbitrary basis can be described by \{$B_{k}=V\pi_{k}V^{\dagger}:k=0,1$\}, where  
$V$ denote the single qubit unitary operator which may be expressed as \cite{luo}
\begin{eqnarray}
V=tI+i\overrightarrow{y}.\overrightarrow{\sigma},~~t^{2}+y_{1}^{2}+y_{2}^{2}+y_{3}^{2}=1
\label{vnm}
\end{eqnarray} 
where $t \in \mathbb{R}$ and $\overrightarrow{y}=(y_{1},y_{2},y_{3}) \in \mathbb{R}^{3} $.\\
Therefore, when Charlie perform measurement $B_{k}$ on his qubit, the three-qubit state $\rho_{ABC}$ projected onto the two-qubit state 
\begin{eqnarray}
\rho_{AB}^{(k)}=\frac{1}{p_{k}}(I\otimes I\otimes B_{k})\rho(I\otimes I\otimes B_{k}^{\dagger}),~~k=0,1
\end{eqnarray} 
where $p_{k}=tr((I\otimes I\otimes B_{k})\rho_{ABC}(I\otimes I\otimes B_{k}^{\dagger}))$ denote the probability of collapsing the three-qubit state to two-qubit state after the measurement performed on the third qubit. The two-qubit state $\rho_{AB}^{(k)}$ shared between Alice and Bob may be used as a resource state when teleporting an arbitrary single qubit state possessed by Alice. We further assume that in the process of single qubit teleportation using a shared two-qubit state, Alice act as a sender and Bob, a receiver. In CQT scheme, the faithfulness of the teleportation may be quantified by the conditioned fidelity denoted by $f_{CT}(\rho_{AB}^{(k)})$ and the non-conditioned fidelity $f_{NC}(\rho_{AB})$.\\
Alternatively, we may also describe the above equivalent situation with the reduced two-qubit state described by the density operator $\rho_{AB}=Tr_{C}(\rho_{ABC})$.  The resulting two-qubit state described by the density operator $\rho_{AB}$ may also be used in transmitting an arbitrary single qubit state through conventional teleportation scheme.\\
If we are not allowing any filtering operation then we may observe that the two-qubit state obtained either through the von-Neumann measurement or through the application of partial trace operation, may or may not be useful as a resource state in quantum teleportation. In this scenario, the controlled teleportation scheme may be helpful in the sense that by controlling the measurement parameter, the controller may be able to increase the teleportation fidelity in the conventional teleportation scheme. Therefore, the enhancement of the teleportation fidelity may be measured by a quantity known as controller's power ($P_{CT}^{(k)}$) of the controlled quantum teleportation. It may be defined as the difference between the conditioned fidelity ($f_{CT}(\rho_{AB}^{(k)})$) and the non-conditined fidelity ($f_{NC}(\rho_{AB})$)
\begin{eqnarray}
	P_{CT}^{(k)}&=&f_{CT}(\rho_{AB}^{(k)})-f_{NC}(\rho_{AB}),~~k=0,1
	\label{def} 
\end{eqnarray}
In the controlled quantum teleportation scheme, there are two basic assumptions: (i) $f_{CT}(\rho_{AB}^{(k)})>\frac{2}{3}$ and (ii) $f_{NC}(\rho_{AB})\leq \frac{2}{3}$. If these two conditions are satisfied by any three-qubit states then we say that the given three-qubit state is useful in CQT scheme.\\
The rest of the paper is organized as follows: In Section-II, we construct an witness operator to detect two-qubit entangled state and then discussed its importance by deriving the lower and upper bound of the expectation value of it with respect to any two-qubit entangled state. In section-III, we estimate the conditioned fidelity and non-conditioned fidelity and thus power of the controller in terms of the expectation value of the witness operator. In secion-IV, we study the controlled teleportation in noisy environment and analyze the relationship between the noisy parameter and the power of the controller. Lastly, we conclude the work.\\

\section{Witness operators}
\noindent An operator $W$ is said to be a witness operator if it satisfies the following two conditions \cite{Guhne}
\begin{eqnarray}
C1.~~ Tr[W\sigma_{sep}]\geq 0, \forall~ \text{separable state}~\sigma_{sep}~~~~~~~~~~~~~~~~~~~~~
\label{c1}\\
C2.~~ Tr[W\sigma_{ent}] < 0, \text{for at least one entangled}~ \text{state}~\sigma_{ent}
\label{c2}
\end{eqnarray}
In this section, our task is to construct witness operator, and study the relationship between the expected value of the constructed witness operator and the Bell-CHSH inequality. We have shown that the constructed witness operator may detect the two-qubit entangled state even when Bell-CHSH inequality unable to detect it. Moreover, we find that the two-qubit entangled states, which are not detected by Bell-CHSH inequality but detected by witness operator, are useful for teleportation.\\
In general, it has already been shown in the literature \cite{horo3} that any two-qubit state described by the density matrix $\rho_{AB}$ violates CHSH inequality if and only if $M(\rho_{AB})>1$. The quantity $M(\rho_{AB})$ may be defined as 
\begin{eqnarray}
	M(\rho_{AB})=u_{1}+u_{2}
	\label{mrho}
\end{eqnarray}
where $u_{1}$ and $u_{2}$ are the two maximum eigenvalues of $T^{\dagger}T$. T denote the $ 3 \otimes 3$ correlation matrix of $\rho_{AB}$ and its entries $t_{ij}$ can be calculated by the formula
\begin{eqnarray}
t_{ij}=Tr[\rho_{AB}(\sigma_{i}\otimes\sigma_{j})],~~ i,j=\{1,2,3\}
\end{eqnarray}
\subsection{Construction of witness operator $W^{(1)}_{ij}$}  
\noindent
 To start with, let us first recall different Bell-CHSH operator defined in $xy-$, $yz-$ and $zx-$ plane, which are collectively denoted as $B^{(ij)}_{CHSH}~(i,j=x,y,z;i\neq j)$ and it is given by \cite{hyllus}
\begin{eqnarray}
B^{(ij)}_{CHSH}&=&\sigma_{i}\otimes \frac{\sigma_{i}+\sigma_{j}}{\sqrt{2}}+ \sigma_{i}\otimes \frac{\sigma_{i}-\sigma_{j}}{\sqrt{2}}\nonumber\\&+&
\sigma_{j}\otimes \frac{\sigma_{i}+\sigma_{j}}{\sqrt{2}}- \sigma_{j}\otimes \frac{\sigma_{i}-\sigma_{j}}{\sqrt{2}}
\label{bchsh}
\end{eqnarray}
Afterward, we will use the short form $B_{ij}$ instead of using the long form $B^{(ij)}_{CHSH}$ throughout the paper. The four Bell states in the computational basis are denoted by $|\phi^{\pm}\rangle$, $|\psi^{\pm}\rangle$ and can be expressed as
\begin{eqnarray}
|\phi^{+}\rangle = \frac{|00\rangle +|11\rangle}{\sqrt{2}}
\label{bellbasis1}\\
|\phi^{-}\rangle = \frac{|00\rangle -|11\rangle}{\sqrt{2}}
\label{bellbasis2}\\
|\psi^{+}\rangle = \frac{|01\rangle +|10\rangle}{\sqrt{2}}
\label{bellbasis3}\\
|\psi^{-}\rangle = \frac{|01\rangle -|10\rangle}{\sqrt{2}}
\label{bellbasis4}
\end{eqnarray}
Now we are in a position to construct the operator $W_{ij}$ that may be expressed in the form as
\begin{equation}
W_{ij}= (\frac{1}{2}+2a)I-A-aB_{ij} ,  i,j=x,y,z~\&~ i\neq j
\label{witdef1}
\end{equation}
where $a$ is a positive real number. The operator $A$  given in Eq. (\ref{witdef1}) may take any form of two-qubit Bell states and other operators have their usual meaning.
In particular, if we take $A=|\phi^{+}\rangle\langle \phi^{+}|$ then the operator $W_{ij}$ reduces to $W^{(\phi^{+})}_{ij}$, where $W^{(\phi^{+})}_{ij}$ is given by
\begin{equation}
W^{(\phi^{+})}_{ij}= (\frac{1}{2}+2a)I-|\phi^{+}\rangle \langle \phi^{+}|-aB_{ij},  i,j=x,y,z~\&~ i\neq j
\label{witdef}
\end{equation}
\textbf{Theorem 1:} The operator $W^{(\phi^{+})}_{ij}$ given in Eq. (\ref{witdef}) is a witness operator.\\
\textbf{Proof:} We call the operator $W^{(\phi^{+})}_{ij}$, a witness operator if it satisfies the conditions $C1$ and $C2$ given in (\ref{c1}) and  (\ref{c2}) respectively.\\
\textbf{(a)} To show the validity of condition $C1$, take the operator $W^{(\phi^{+})}_{ij}$ and consider an arbitrary two-qubit separable state described by the density operator $\sigma_{sep}$. The expectation value of the operator $W^{(\phi^{+})}_{ij}$ with respect to  $\sigma_{sep}$ is given by 
\begin{equation}
 Tr[W^{(\phi^{+})}_{ij}\sigma_{sep}]= (\frac{1}{2}+2a)-\langle
 \phi^{+}|\sigma_{sep}|\phi^{+}\rangle -aTr[B_{ij}\sigma_{sep}]
\label{expecval1}
 \end{equation}
If $F(\sigma_{sep})$ denote the singlet fraction \cite{Bennett} of the state $\sigma_{sep}$ then we have 
\begin{eqnarray}
F(\sigma_{sep}) \geq \langle \phi^{+}|\sigma_{sep}|\phi^{+}\rangle
\label{singfrac}
\end{eqnarray}
Using Eq. (\ref{singfrac}) in Eq. (\ref{expecval1}), we get
\begin{eqnarray}
Tr[W^{(\phi^{+})}_{ij}\sigma_{sep}] &\geq& (\frac{1}{2}+2a)-F(\sigma_{sep}) -aTr[B_{ij}\sigma_{sep}]\nonumber\\
\label{expecval}
\end{eqnarray}
For any separable state $\sigma_{sep}$, we have $-2\leq Tr[B_{ij}\sigma_{sep}] \leq 2$. Thus, for $a>0$, the inequality (\ref{expecval}) reduces to
\begin{equation}
Tr[W^{(\phi^{+})}_{ij}\sigma_{sep}] \geq \begin{cases}
\frac{1}{2}-F(\sigma_{sep})+2a, & Tr[B_{ij}\sigma_{sep}]\in [-2,0]\\
\frac{1}{2}-F(\sigma_{sep}), & Tr[B_{ij}\sigma_{sep}]\in [0,2]	\end{cases}
\end{equation}
Since the singlet fraction of any separable state $\sigma_{sep}$ satisfies the inequality $F(\sigma_{sep}) \leq  \frac{1}{2}$ so for any $a>0$ and for any separable state $\sigma_{sep}$, we have $Tr[W^{(\phi^{+})}_{ij}\sigma_{sep}] \geq 0$. Hence $C1$ is verified.\\
\textbf{(b)} To prove the validity of the condition $C2$, it is enough to show that there exist an entangled state $\sigma_{ent}$ for which $Tr[W^{(\phi^{+})}_{ij}\sigma_{ent}]<0$. For this, let us consider an entangled state $\sigma_{ent}$, which may be defined as \cite{Mhorodecki}
\begin{eqnarray}
\sigma_{ent}&=& p|\phi^{+}\rangle \langle \phi^{+}|+ \frac{1-p}{4}I , \frac{1}{3}<p \leq 1 
\label{state}
\end{eqnarray}
Let us now consider the operator $B_{yz}$ which is defined in the interval $\frac{1}{3}<p \leq 1$. In this interval, we find that the state $\sigma_{ent}$ satisfy the Bell-CHSH inequality as it is clearly evident from the equation given below
\begin{eqnarray}
Tr[B_{yz}\sigma_{ent}]= 0
\label{byz}
\end{eqnarray}
The operator $W^{(\phi^{+})}_{yz}$ may be expressed as 
\begin{equation}
W^{(\phi^{+})}_{yz}= (\frac{1}{2}+2a)I-|\phi^{+}\rangle \langle \phi^{+}|-aB_{yz},~~a>0
\label{witdefyz}
\end{equation}
The expectation value of the operator $W^{(\phi^{+})}_{yz}$ with respect to the state $\sigma_{ent}$ is given by 
\begin{eqnarray}
Tr[W^{(\phi^{+})}_{yz}\sigma_{ent}]&=& \frac{1}{2}+ 2a - \langle \phi^{+}|\sigma_{ent}|\phi^{+}\rangle -aTr[B_{yz}\sigma_{ent}]\nonumber \\
&=& \frac{1}{2}+ 2a -\frac{1+3p}{4}\nonumber \\
&=& \frac{1-3p}{4}+ 2a \nonumber \\
&<&  0,~ a \in [0,0.001],~\frac{1}{3}<p\leq 1 \nonumber
\label{expecval3}
\end{eqnarray} 
Thus, there exist an entangled state $\sigma_{ent}$ for which $Tr[W^{(\phi^{+})}_{yz}\sigma_{ent}]<0$. Therefore, $C2$ is verified.\\
Thus, we can now say that the operator $W^{(\phi^{+})}_{yz}$ may serve as a valid entanglement witness operator. Similarly, it can be shown that there exist a range of the parameter $a$ for which $Tr[W^{(\phi^{+})}_{xy}\sigma_{ent}]<0$ and  $Tr[W^{(\phi^{+})}_{zx}\sigma_{ent}]<0$. Hence, the witness operator $W^{(\phi^{+})}_{ij}$ for any $i,j=x,y,z;i\neq j$ is a witness operator.\\
Moreover, if we replace the operator $A$ by other Bell states like $|\phi^{-}\rangle\langle \phi^{-}|$ or $|\psi^{\pm}\rangle\langle \psi^{\pm}|$ then it can be shown that the corresponding operators $W^{(\phi^{-})}_{ij}$ or $W^{(\psi^{\pm})}_{ij}$ may serve as witness operator for any $i,j=x,y,z;i\neq j$. Therefore, the witness operators $W^{(\phi^{-})}_{ij}$, $W^{(\psi^{\pm})}_{ij}$ may be expressed in the following way:\\
\begin{equation}
W^{(\phi^{-})}_{ij}= (\frac{1}{2}+2a)I-|\phi^{-}\rangle \langle \phi^{-}|-aB_{ij} ,  i,j=x,y,z~~ i\neq j 
\label{witdef11}
\end{equation}
\begin{equation}
W^{(\psi^{\pm})}_{ij}= (\frac{1}{2}+2a)I-|\psi^{\pm}\rangle \langle \psi^{\pm}|-aB_{ij} ,  i,j=x,y,z~~ i\neq j 
\label{witdef2}
\end{equation}

\subsection{Characteristic of the introduced witness operator}
\noindent In this section, we may take into account the Bell state $|\phi^{+}\rangle$ and then discuss the relation between the three quantities such as (i) $M(\rho)$, which determine whether the quantum state violating the Bell-CHSH inequality (ii) $F(\rho)$ denoting the singlet fraction of the state $\rho$ that determine whether the state is useful as a resource in quantum teleportation \cite{horo4} and (iii) the expectation value of the witness operator $W^{(\phi^{+})}_{ij}$ that detect the signature of the entanglement. Specifically, we derived here the lower and upper bound of the the expectation value of the witness operator $W^{(\phi^{+})}_{ij}$. Using these bounds, we have obtained few results that focusses on the condition for which the witness operator may detect or may not detect the entangled state. Furthermore, we note that all the results obtained by considering the operator $|\phi^{+}\rangle\langle \phi^{+}|$ may also be obtained by considering the other three Bell operators such as $|\phi^{-}\rangle\langle \phi^{-}|$, $|\psi^{\pm}\rangle\langle \psi^{\pm}|$.\\ 
\textbf{Result-1:} Consider an entangled state $\rho_{ent}$ such that $M(\rho_{ent})\leq 1$. Then the lower and upper bound of the expectation value of the witness operator $W^{(\phi^{+})}_{ij}$ with respect to an entangled state $\rho_{ent}$ is given by
\begin{eqnarray}
U(a) \leq	Tr[W^{(\phi^{+})}_{ij}\rho_{ent}] \leq L(a) ,~~a>0
\label{p1}
\end{eqnarray}
where $U(a)=(\frac{1}{2}-F(\rho_{ent}))+2a(1-\sqrt{M(\rho_{ent})})$ and $L(a)=\frac{1}{2}- \langle \phi^{+}|\rho_{ent}|\phi^{+}\rangle +4a$.\\
The proof is given in the Appendix-A.\\
The inequality (\ref{p1}) estimates the lower and upper bound of  the expectation value of the witness operator $W^{(\phi^{+})}_{ij}$ with respect to any two-qubit entangled state. Further, we note that the lower bound of $Tr[W^{(\phi^{+})}_{ij}\rho_{ent}]$ depends on two inequalities such as (i) the singlet fraction $F(\rho_{ent})$ and (ii) a quantity $M(\rho_{ent})$. Based on these two quantities, 
we can make the following observations from the inequality (\ref{p1}):\\
\textbf{Observation-1:} If there exist any two-qubit entangled state $\rho_{ent}$ such that $M(\rho_{ent})\leq 1$ and $F(\rho_{ent})\leq \frac{1}{2}$ then it is clear from Eq. (\ref{p1}) that the witness operator $W^{(\phi^{+})}_{ij}$ cannot detect the entangled state $\rho_{ent}$.\\
This observation may be illustrated by the following example: Let us consider the two-qubit state \cite{versatrate}
\begin{eqnarray}
	\rho_{F}&=& F|\phi^{+}\rangle \langle \phi^{+}|+ (1-F)|01\rangle \langle 01| , ~~\frac{1}{3}<F \leq \frac{1}{2} \nonumber \\
\end{eqnarray}
where $F$ denotes the singlet fraction of the state. One may easily verify that the state $\rho_{F}$ is an entangled state when $\frac{1}{3}<F \leq \frac{1}{2}$.\\
The expectation value of the Bell operators $B_{xy}$, $B_{yz}$ and $B_{xz}$ in different setting with respect to the state $\rho_{F}$ is given by
\begin{eqnarray}
	\langle B_{xy} \rangle_{\rho_{F}}=0,
	\label{f1}\\
	\langle B_{yz} \rangle_{\rho_{F}}\in (-0.9428,-0.7071), 
	\label{f21}\\
	\langle B_{xz} \rangle_{\rho_{F}} \in (0,0.707107)
	\label{f31}
\end{eqnarray}
Therefore, using using (\ref{f1}), (\ref{f21}) and (\ref{f31}), we can find that the state $\rho_{F}$ satisfies the Bell-CHSH inequality. Let us now calculate the expectation value of the corresponding witness operators $W^{(\phi^{+})}_{xy}$, $W^{(\phi^{+})}_{yz}$, and $W^{(\phi^{+})}_{zx}$ with respect to the state $\rho_{F}$. For positive 'a', the expectation values are given by 
\begin{eqnarray}
	Tr[W^{(\phi^{+})}_{xy}\rho_{F}]&=& \frac{1}{2}+2a-F>0 
	\label{wx1}\\
	Tr[W^{(\phi^{+})}_{yz}\rho_{F}]&=& (\frac{1}{2}-F)+a[2+\sqrt{2}(1-F)]>0 \nonumber\\ 
	\label{wy2}\\
	Tr[W^{(\phi^{+})}_{zx}\rho_{F}]&=& (\frac{1}{2}-F)+a[2+\sqrt{2}(1-3F)]>0 \nonumber\\ 
	\label{wz3}
\end{eqnarray}
Thus, it is clear from (\ref{wx1}), (\ref{wy2}) and (\ref{wz3}) that the entangled state $\rho_{F}$,
is not detected by the witness operator $W^{(\phi^{+})}_{xy}$. The observation-1 is now verified for a particular quantum state described by the density operator $\rho_{F}$. But, in general, from the inequality (\ref{p1})  we can conclude that if any quantum entangled state $\rho$ satisfies $M(\rho)\leq 1$ and $F(\rho)\leq \frac{1}{2}$ then the witness operator $W^{(\phi^{+})}_{ij},~i,j=x,y,z,~i\neq j $ does not detect the entangled state $\rho$.\\
\textbf{Observation-2:} If there exist any two-qubit entangled state $\rho_{ent}$ which is useful in teleportation i.e. $F(\rho_{ent})> \frac{1}{2}$ then the witness operator $W^{(\phi^{+})}_{ij}$ may detect the entangled state $\rho_{ent}$ when the parameter $a$ lies in some specific range. This observation may be written in the form of another result, which is stated below:\\
\textbf{Result-2:} Let us consider a two-qubit entangled state described by a density operator $\rho_{ent}$. If $F(\rho_{ent}) > \frac{1}{2}$
and if the parameter $a$ lies in the range $0<a\leq \frac{\langle \phi^{+}|\rho_{ent}| \phi^{+}\rangle-\frac{1}{2}}{4}$ then the witness operator $W_{ij}^{(\phi^{+})}$ detect the entangled state $\rho_{ent}$. \\
\textbf{Proof:} The expectation value of the witness operator $W_{ij}^{(\phi^{+})}$ $(i,j= x,y,z; i\neq j)$ with respect to the entangled state $\rho_{ent}$ can be written as
\begin{eqnarray}
Tr[W_{ij}^{(\phi^{+})}\rho_{ent}] &=& \frac{1}{2}- \langle \phi^{+}|\rho_{ent}| \phi^{+}\rangle +a\times\nonumber\\&&(2-Tr[B_{ij}\rho_{ent}]) 
\label{w1}
\end{eqnarray}
Using Eq. (\ref{p1}), it can be easily shown that if $F(\rho_{ent})>\frac{1}{2}$ and if whether the state $\rho_{ent}$ satisfies the Bell-CHSH inequality or violate it, the upper and lower bound of $Tr[W_{ij}^{(\phi^{+})}\rho_{ent}]$ will be a negative quantity. Thus, we have
\begin{eqnarray}
Tr[W_{ij}^{(\phi^{+})}\rho_{ent}]= \text{a negative quantity}
\label{w1neg}	
\end{eqnarray}
Therefore, (\ref{w1neg}) clearly indicate the fact that the witness operator $W_{ij}^{(\phi^{+})}$  $(i,j= x,y,z; i\neq j)$ detect the entangled state $\rho_{ent}$. Hence proved the result.\\
We will now verify Result-2 by considering the Bell state $|\psi^{-}\rangle$ instead of taking $|\phi^{+}\rangle$. To verify Result-2, let us consider the two-qubit state described by the density operator $\rho(\theta)$
\begin{eqnarray}
	\rho(\theta)= \frac{1}{2}\begin{pmatrix} 
		a(\theta) & 0& 0& 0\\
		0 & b(\theta) & c(\theta) & 0\\
		0 & c(\theta) & d(\theta) & 0\\
		0 & 0& 0& e(\theta) \\
	\end{pmatrix},~ 0\leq \theta \leq 0.4175\pi \nonumber \\
\end{eqnarray}
where $a(\theta)=(3-2\sqrt{2})Sin^{2}\theta $, $b(\theta)=(3-2\sqrt{2})Cos^{2}\theta$, $c(\theta)=(1-\sqrt{2})Cos\theta$, $d(\theta)=1+(2\sqrt{2}-2)Sin^{2}\theta$ and $e(\theta)=(2\sqrt{2}-2)Cos^{2}\theta$.\\ 
It can be easily verified that $\rho(\theta)$ is an entangled state and $M(\rho(\theta))<1$  for $\theta \in [0,0.4175\pi]$. Thus, the entangled state $\rho(\theta)$ will satisfy Bell-CHSH inequality for $\theta \in [0,0.4175\pi]$ and thus it is undetected by Bell-CHSH operator. Further, the singlet fraction of $\rho(\theta)$, i.e., $F(\rho(\theta))$ can be calculated as 
\begin{eqnarray}
F(\rho(\theta))= \frac{1}{8}(3 + 4 (-1 + \sqrt{2}) Cos\theta+ (5 - 4\sqrt{2}) Cos(2\theta)) \nonumber
\end{eqnarray}
We can verify that $F(\rho(\theta))> \frac{1}{2}$ when $\theta \in [0, 0.4175\pi]$ and $a \in (0,0.00560188]$.\\ 
By direct calculation, we obtain the value of the following expressions in terms of the state parameter $\theta$ as
\begin{eqnarray}
\langle B_{xy} \rangle_{\rho(\theta)}&=&2 (-2 + \sqrt{2}) Cos\theta
\nonumber \\
\langle B_{yz} \rangle_{\rho(\theta)}=\langle B_{xz} \rangle_{\rho(\theta)}&=& \frac{1}{\sqrt{2}}(-1 - 2 (-1 + \sqrt{2}) Cos\theta \nonumber \\ &+&  (-5 + 4\sqrt{2} ) Cos2\theta)  \nonumber\\
Tr[\rho(\theta) |\psi^{-} \rangle\langle \psi^{-}|]&=& \frac{1}{8}(3 + 4 (-1 + \sqrt{2}) Cos\theta \nonumber \\ &+& (5 - 4\sqrt{2}) Cos(2\theta)) 
\end{eqnarray}
We are now in a position to calculate the expectation value of the witness operatr $W_{ij}^{(\psi^{-})}$ with respect to the state $\rho(\theta)$. It is given by 
\begin{eqnarray}
	Tr[W_{xy}^{(\psi^{-})}\rho(\theta)]&=& \frac{1}{8} (1 + 16 a - 
	4 (-1 + \sqrt{2} + 4 (-2 \nonumber \\ &+& \sqrt{2}) a) Cos\theta + (-5 + 4\sqrt{2}) Cos2\theta) \nonumber \\
	Tr[W_{yz}^{(\psi^{-})}\rho(\theta)]&=& \frac{1}{8} (1 + 16a + 4 \sqrt{2} a -
	4 (-1 + \sqrt{2} +\nonumber \\ &&  2 (-2 + \sqrt{2}) a) Cos\theta +(-5 + 4 \sqrt{2}+ \nonumber \\&& 4 (-8 + 5\sqrt{2}) a) Cos2\theta) \nonumber\\
Tr[W_{xz}^{(\psi^{-})}\rho(\theta)]&=& Tr[W_{yz}^{(\psi^{-})}\rho(\theta)] 
\end{eqnarray}
We find that the witness operator $W_{xy}^{(\psi^{-})}$ detect the state $\rho(\theta)$ when $a\in (0,0.00032]$ \& $\theta \in [0,0.4175\pi]$. We also find that witness operator $W_{yz}^{(\psi^{-})}$ \& $W_{xz}^{(\psi^{-})}$ detect the state $\rho(\theta)$ when $a\in (0,0.00016]$ \& $0\leq \theta < 0.4175\pi $. Therefore, there exist a range of the parameter $a$ for which the entangled state $\rho(\theta)$ is detected by the witness operator $W_{ij}^{(\psi^{-})}$ when $i,j=x,y,z;i\neq j$.\\
Now, we are in a position to derive the non-trivial lower bound of the teleportation fidelity when $\rho_{ent}$ is used as a resource state in quantum teleportation. It may be expressed in terms of the expectation value of the witness operator and M($\rho_{ent}$),\\
\textbf{Result-3:} If there exist an entangled state described by the density operator $\rho_{ent}$, which satisfies the Bell-CHSH inequality but detected by the witness operator $W^{(\phi^{+})}_{ij}$, then the entangled state $\rho_{ent}$ is useful in teleportation with teleportation fidelity $f(\rho_{ent})$, which satisfies the inequality
\begin{eqnarray}
f(\rho_{ent}) \geq	\frac{2}{3}\{1-Tr[W^{(\phi^{+})}_{ij}\rho_{ent}]+2a(1-\sqrt{M(\rho_{ent})})\}\nonumber\\
\label{fid1}
\end{eqnarray} 
where $a\in (0,\frac{\frac{1}{2}+Tr[W^{(\phi^{+})}_{ij}\rho_{ent}]}{2(1-\sqrt{M(\rho_{ent})})}]$.\\
The proof of Result-3 is given in Appendix-B.
\section{Estimation of controller's power} 
\noindent We have assumed here that the controlled teleportation scheme involve three parties namely Alice (A), Bob (B) and Charlie (C), who have shared a three-qubit state. In this protocol, the measurement is performed by Charlie (acting as a controller) on his qubit. As a result of the measurement, the two-qubit state will be shared between Alice and Bob described by the density operator $\rho_{AB}$. The shared state $\rho_{AB}$ may or may not violate the Bell-CHSH inequality and accordingly the state may or may not be useful in the conventional teleportation scheme \cite{horo4}. Therefore, the study of the violation of Bell-CHSH inequality is important in this scenario and thus we consider it here as the CHSH game \cite{jon}. In the CHSH game, we assume that the two distant players, Alice (A) and Bob (B) receive binary questions $s, t \in \{0, 1\}$ respectively, and similarly their answers $a, b \in \{0, 1\}$ are single bits. Alice
and Bob win the CHSH game if their answers satisfy $a \oplus b = s ·t$. Thus, CHSH game can be considered as a particular example of XOR games. In this game, the non-locality of the shared state $\rho_{AB}$ may be determined when Alice and Bob perform
measurements on their respective qubit and the outcomes
of their measurements are correlated. Therefore, the maximum probability $P_{ij}$ of winning the game overall strategy is given by \cite{jon} 
\begin{eqnarray}
P_{ij}&=&\frac{1}{2}[1+\frac{\langle B_{ij}\rangle _{\rho_{AB}}}{4}]
\label{pmax}
\end{eqnarray}
where $\langle B_{ij}\rangle _{\rho_{AB}}={Tr\rm}[(A_{0}\otimes B_{0}+A_{0}\otimes B_{1}+A_{1}\otimes B_{0}-A_{1}\otimes B_{1})\rho_{AB}]$. Since the maximum probability of winning the game depends on the expectation value of the Bell operator $B_{ij}$, so $P^{max}$ is somehow related to the non-locality of the state $\rho_{AB}$. Adding the known fact that the state $\rho_{AB}$ violate the Bell-CHSH inequality if $\langle B_{ij}\rangle _{\rho_{AB}}>2$ and thus, we find that the state $\rho_{AB}$ is non-local when $P^{max}>\frac{3}{4}$. Hence, the shared state $\rho_{AB}$ may be useful for teleportation when $P^{max}>\frac{3}{4}$.\\
It may be easily shown that the winning probability $P_{ij}$ may be estimated in terms of the expectation value of the witness operator $W_{ij}$ with respect to the state $\rho_{AB}$. The result may be stated as\\
\textbf{Lemma:} The probability $P_{ij}$ of the CHSH game may be estimated as\\
(i) When $F(\rho_{AB})\leq \frac{1}{2}$
\begin{eqnarray}
\frac{3}{4}-\frac{Tr[W_{ij}\rho_{AB}]}{8a} \leq P_{ij} \leq 1
\label{pmaxest1}
\end{eqnarray}
(i) When $F(\rho_{AB})> \frac{1}{2}$
\begin{eqnarray}
0\leq P_{ij} < \frac{3}{4}-\frac{Tr[W_{ij}\rho_{AB}]}{8a}
\label{pmaxest2}
\end{eqnarray}
where $F(\rho_{AB})$ denote the singlet fraction of the state $\rho_{AB}$.\\
\textbf{Proof:} Let us recall the witness operator $W_{ij}^{(\phi^{+})}$ given in (\ref{witdef}). Therefore, Using (\ref{witdef}) and ($\ref{pmax}$), the expression for $P_{ij}$ may be re-written as
\begin{eqnarray}
P_{ij}= \frac{3}{4}+\frac{1}{8a}[\frac{1}{2}- \langle\phi^{+}|\rho_{AB}|\phi^{+}\rangle-  Tr[W_{ij}^{(\phi^{+})}\rho_{AB}]] \nonumber \\
\label{pmaxexp}
\end{eqnarray}
Using the fact that $\langle\phi^{+}|\rho_{AB}|\phi^{+}\rangle \leq F(\rho_{AB})$ and considering the two different cases i) $F(\rho_{AB}) \leq \frac{1}{2}$ and ii) $F(\rho_{AB})> \frac{1}{2}$ separately, we can easily obtain the above estimation given in (\ref{pmaxest1}) and (\ref{pmaxest2}). The above result may be proved in the same way for $W_{ij}^{(\phi^{-})}$, $W_{ij}^{(\psi^{+})}$ and $W_{ij}^{(\psi^{-})}$. Hence proved.\\
In this section, we aim to derive the lower and upper bound of the controller's power using the winning probability of the CHSH game. We found that the controller's power ($P_{CT}^{(k)}$) is always upper bounded by $\frac{1}{2}$ while the lower bound may be estimated in terms of a quantity $M(\rho^{(k)}_{AB})$ and the expectation value of the witness operator $W_{ij}^{(\phi^{+})}$ with respect to the state $\rho^{(k)}_{AB}$. The state $\rho^{(k)}_{AB}$ has been obtained when charlie perform measurement $B_{k}$ on his qubit.\\
\subsection{Estimation of non-conditioned teleportation fidelity}
\noindent Let us suppose that the three-qubit state shared between Alice (A), Bob (B) and Charlie (C) is described by the density operator $\rho_{ABC}$. The reduced two-qubit state shared between Alice and Bob is described by the density operator $\rho_{AB}=Tr_{C}(\rho_{ABC})$. If $\rho_{AB}$ is used as a resource state in quantum teleportation then the faithfulness of the teleportation is determined by the non-conditioned teleportation fidelity which is denoted by $f_{NC}(\rho_{AB})$. 
The non-conditioned fidelity can be expressed in terms of the correlation tensor $T_{AB}$ as \cite{horo4} 
\begin{eqnarray}
f_{NC}(\rho_{AB})= \frac{3+||T_{AB}||_{1}}{6}
\label{fnctensor}
\end{eqnarray}
where $||.||_{1}$ denote the trace norm.\\
To express $f_{NC}(\rho_{AB})$ in terms of witness operator, we recall the expression of $P_{ij}$ given in (\ref{pmaxexp}). It may be re-written as
\begin{eqnarray}
P_{ij}= \frac{3}{4}+\frac{1}{8a}(\frac{1}{2}- \langle\phi^{+}|\rho_{AB}|\phi^{+}\rangle-  Tr[W_{ij}^{(\phi^{+})}\rho_{AB}])\nonumber
\end{eqnarray}
Using $\langle\phi^{+}|\rho_{AB}|\phi^{+}\rangle \leq F(\rho_{AB})$ in the expression of $P_{ij}$, we get the inequality as
\begin{eqnarray}
Tr[W_{ij}^{(\phi^{+})}\rho_{AB}]\geq 8a(\frac{3}{4}-P_{ij})+\frac{1}{2}-F(\rho_{AB})
\end{eqnarray}
One of the assumption to execute the controlled quantum teleportation scheme is that the non-conditioned teleportation fidelity must be less than $\frac{2}{3}$. Thus, considering $F(\rho_{AB})\leq \frac{1}{2}$ and $P_{ij}\leq \frac{3}{4}$, we get 
\begin{eqnarray}
\frac{1}{2}-Tr[W_{ij}^{(\phi^{+})}\rho_{AB}] \leq	F(\rho_{AB})\leq \frac{1}{2}
\label{fnc}
\end{eqnarray}
Using the relation between singlet fraction $(F(\rho_{AB}))$ and non-conditioned teleportation fidelity $(f_{NC}(\rho_{AB}))$, the inequality (\ref{fnc}) may be expressed in terms of $(f_{NC}(\rho_{AB}))$. Therefore, the inequality (\ref{fnc}) may be re-expressed as 
\begin{eqnarray}
\frac{2}{3}(1-Tr[W_{ij}^{(\phi^{+})}\rho_{AB}])\leq f_{NC}(\rho_{AB})\leq \frac{2}{3}
\label{fnc1}
\end{eqnarray} 
While constructing the witness operator $W_{ij}^{(\phi^{+})}$, we should be careful in choosing the positive value of the parameter $a$. The value of $a$ is choosen in such a way that $Tr[W_{ij}^{(\phi^{+})}\rho_{AB}]\geq 0$.
\subsection{Estimation of the conditioned teleportation fidelity}
\noindent In the controlled teleportation protocol, when the controller charlie measures on his qubit, the three-qubit state $\rho_{ABC}$ reduces to $\rho_{AB}^{(k)}$ according to the measurement outcome $k=0,1$. If Alice and Bob uses the shared state $\rho_{AB}^{(k)}$ as a resource state in the teleportation protocol then the fidelity of the teleportation may be termed as conditioned teleportation fidelity and it is denoted by $f_{C}(\rho_{AB}^{(k)})$. There is an interesting relationship between the conditioned teleportation fidelity and the partial tangle $\tau_{AB}$ and it is given by \cite{lee}
\begin{eqnarray}
f_{C}(\rho_{AB}^{(k)})=\frac{2+\tau_{AB}^{(k)}}{3}, k=0,1
\label{confidtau}
\end{eqnarray} 
To implement the controlled quantum teleportation, it is assumed that $f_{C}(\rho_{AB}^{(k)})>\frac{2}{3}$.
Therefore, the conditioned teleportation fidelity $f_{C}(\rho_{AB}^{(k)})$ may be estimated by using the $Result-3$
\begin{eqnarray}
L_{C} \leq f_{C}(\rho_{AB}^{(k)}) 
\label{fid12}
\end{eqnarray} 
where $L_{C}=\frac{2}{3}\{1-Tr[W^{(\phi^{+})}_{ij}\rho_{AB}^{(k)}]+2a(1-\sqrt{M(\rho_{AB}^{(k)})})\}$.\\
The condition of controlled teleportation will be met when the witness operator detect the entangled state $\rho_{AB}^{(k)}$ and also when the entangled state $\rho_{AB}^{(k)}$ satisfies the Bell-CHSH inequality. The value of the parameter $a$ involved in the witness operator will be chosen in such a way that the witness operator detect $\rho_{AB}^{(k)}$.    

\subsection{Lower and Upper bound of the controller's power}
\noindent The power of the controlled quantum teleportation for the $k^{th}$ measurement outcome may be defined as
\begin{eqnarray}
P_{CT}^{(k)}= f_{C}(\rho_{AB}^{(k)})- f_{NC}(\rho_{AB}), k=0,1
\label{power1}
\end{eqnarray}
Using (\ref{fnctensor}) and (\ref{confidtau}), the expression of the power given in (\ref{power1}) reduces to
\begin{eqnarray}
P_{CT}^{(k)}=\frac{1}{6}+\frac{1}{6}(2\tau_{AB}^{(k)}-\lVert T_{AB}\rVert_{1})
\label{power2} 
\end{eqnarray}
Our task is now to estimate the value of $||T_{AB}||_{1}$ and $\tau_{AB}^{(k)}$.\\
\textbf{(i) Estimation of $||T_{AB}||_{1}$:} Let us recall (\ref{fnc1}) and using (\ref{fnctensor}) in it, we get the estimation of $||T_{AB}||_{1}$ which is given by
\begin{eqnarray}
1-4Tr[W_{ij}^{NC}\rho_{AB}]\leq ||T_{AB}||_{1} \leq 1
\label{esttensor} 
\end{eqnarray}
In this case, the parameter $a$ is chosen in such a way that the witness operator $W_{ij}^{NC}$ does not detect the state $\rho_{AB}$. Therefore, we can put the restriction on $Tr[W_{ij}^{NC}\rho_{AB}]$ as
\begin{eqnarray}
0\leq Tr[W_{ij}^{NC}\rho_{AB}]\leq \frac{1}{4}
\label{restriction} 
\end{eqnarray} 
The upper bound of $Tr[W_{ij}^{NC}\rho_{AB}]$ is obtained by using the condition $||T_{AB}||_{1}\geq 0$.\\
\textbf{(ii) Estimation of $\tau_{AB}^{(k)}$:} Using (\ref{confidtau}) and (\ref{fid12}) and simplifying the inequality, we get
\begin{eqnarray}
4a(1-\sqrt{M(\rho_{AB}^{(k)})})-2Tr[W_{ij}^{C}\rho_{AB}^{(k)}] \leq \tau_{AB}^{(k)} \leq 1
\label{esttau} 
\end{eqnarray} 
The parameter $a$ in the LHS of above inequality (\ref{esttau}) can be chosen in such a way that the witness operator $W_{ij}^{C}$ detect the state $\rho_{AB}^{k}$.\\
Now, we are in a position to derive the lower and upper bound of the power $P_{CT}^{(k)}$. To start with, let us
use the upper bound of $||T_{AB}||_{1}$ and the lower bound of $\tau_{AB}^{(k)}$ in the expression (\ref{power2}) of the power of the controlled teleportation. Therefore, it reduces the power given in (\ref{power2}) to the inequality that gives the lower bound as 
\begin{eqnarray}
\frac{4a}{3} (1-\sqrt{M(\rho_{AB}^{(k)})})-\frac{2}{3}Tr[W_{ij}^{(C)}\rho_{AB}^{(k)}] \leq P_{CT}^{(k)}
\label{plb}
\end{eqnarray}
Similarly, using the lower bound of $||T_{AB}||_{1}$ and the upper bound of $\tau_{AB}^{(k)}$ in the expression of the power of the controlled teleportation, we get the upper bound of the power which is given by 
\begin{eqnarray}
P_{CT}^{(k)} \leq \frac{1}{3}+\frac{2}{3}Tr[W_{ij}^{NC}\rho_{AB}]
\label{power3}
\end{eqnarray} 
Further, if we use the restriction given in (\ref{restriction}) then the inequality (\ref{power3}) reduces to 
\begin{eqnarray}
P_{CT}^{(k)} \leq \frac{1}{2}
\label{pub}
\end{eqnarray} 
Combining (\ref{plb}) and (\ref{pub}), we get 
\begin{eqnarray}
\frac{4a}{3} (1-\sqrt{M(\rho_{AB}^{(k)})})-\frac{2}{3}Tr[W_{ij}^{(C)}\rho_{AB}^{(k)}] \nonumber \\ \leq P_{CT}^{(k)} \leq \frac{1}{2}
\label{ublb}
\end{eqnarray}
\subsection{Estimation of the lower bound of the power for pure three-qubit states}
\noindent In this section, we study the controlled quantum teleportation protocol by considering the pure three-qubit states such as standard GHZ state, Maximally Slice State (MSS) and a W class of states. Then we estimate the lower bound of the power of the controller for all the above mentioned states.\\
\noindent Let us consider a three-qubit standard GHZ state of the form 
\begin{eqnarray}
|\psi^{(1)}\rangle_{CAB}&=& \lambda_{0}|000\rangle +\lambda_{4}|111\rangle,~~\lambda_{0}^{2}+\lambda_{4}^{2}=1  
\label{ghzstate}
\end{eqnarray}
Now, to execute the controlled teleportation scheme with the three-qubit state described by the density operator $\rho_{CAB}^{(1)}= |\psi^{(1)}\rangle_{CAB}\langle\psi^{(1)}|$, the assumptions on the non-conditioned fidelity and conditioned fidelity must be fulfilled. Therefore, we need to calculate the non-conditioned fidelity and conditioned fidelity and thus the power of the controller.\\
(i) Non-conditioned fidelity: We trace out system C from the three-qubit state $\rho_{CAB}^{(1)}$. The resulting two qubit state $\rho_{AB}^{(1)}$ is given by
\begin{eqnarray}
\rho_{AB}^{(1)}=Tr_{C}(\rho_{CAB}^{(1)})=\lambda_{0}^{2}|00\rangle \langle00| + \lambda_{4}^{2}|11\rangle \langle11|
\end{eqnarray}
Using $\rho_{AB}^{(1)}$ as a resource state in quantum teleportation, the non-conditioned fidelity can be calculated as
\begin{eqnarray}
f_{NC}(\rho_{AB}^{(1)})=\frac{2}{3}
\label{ncfghz}
\end{eqnarray}
(ii) Conditioned fidelity: Charlie, performed measurement on his qubit in the  single qubit generalised basis $\{B_{0},B_{1}\}$. After the measurement, the state collapses either to $\rho_{AB}^{G(0)}$ or  $\rho_{AB}^{G(1)}$, where  
\begin{eqnarray}
\rho_{AB}^{G(0)}&=&\frac{1}{p_{0}}\bigg( (t^{2}+y_{3}^{2})\lambda_{0}^{2}|00\rangle \langle00|+ \lambda_{0}\lambda_{4}(-ty_{2}+y_{1}y_{3}\nonumber \\ &+&\iota(ty_{1}+y_{2}y_{3}))|00\rangle \langle11| + \lambda_{0}\lambda_{4}(-ty_{2}+y_{1}y_{3}\nonumber \\ &-&\iota(ty_{1}+y_{2}y_{3}))|11\rangle \langle00|+ \lambda_{4}^{2}(y_{1}^{2}+y_{2}^{2}) |11\rangle \langle11|\bigg), \nonumber \\&&
\text{where}~~~~ p_{0}=(t^{2}+y_{3}^{2})\lambda_{0}^{2}+(y_{1}^{2}+y_{2}^{2})\lambda_{4}^{2}
\end{eqnarray}
\begin{eqnarray}
\rho_{AB}^{G(1)}&=&\frac{1}{p_{1}}\bigg( (y_{1}^{2}+y_{2}^{2})\lambda_{0}^{2}|00\rangle \langle00|+ \lambda_{0}\lambda_{4}(ty_{2}-y_{1}y_{3}\nonumber \\ &-&i(ty_{1}+y_{2}y_{3}))|00\rangle \langle11| + \lambda_{0}\lambda_{4}(ty_{2}-y_{1}y_{3}\nonumber \\ &+&i(ty_{1}+y_{2}y_{3}))|11\rangle \langle00|+ \lambda_{4}^{2}(t^{2}+y_{3}^{2}) |11\rangle \langle11|\bigg), \nonumber \\&&
\text{where}~~~~p_{1}=(y_{1}^{2}+y_{2}^{2})\lambda_{0}^{2}+(y_{3}^{2}+t^{2})\lambda_{4}^{2}
\end{eqnarray}
Let us now use the state $\rho_{AB}^{G(0)}$ for the teleportation of a single qubit. We choose the normalized measurement parameters $(y_{1},y_{2},y_{3},t)$ in such a way that the conditioned fidelity is greater than $\frac{2}{3}$ i.e. $f_{C}(\rho_{AB}^{G(0)})>\frac{2}{3}$ and also the normalization condition (\ref{vnm}) holds. Therefore, choosing the measurement parameters $y_{1}=-0.25$, $y_{2}=-0.49$, $y_{3}=0.39$ and $t=-0.74$, we can calculate the conditioned fidelity in terms of the state parameter $\lambda_{4}$ as 
\begin{eqnarray}
f_{C}(\rho_{AB}^{G(0)})&=& \frac{1}{1.79958-\lambda_{4}^{2}}(1.19972 - 0.666667\lambda_{4}^{2} \nonumber\\ &+& 0.799676\lambda_{4}\sqrt{1-\lambda_{4}^{2}})
\label{cfghz}
\end{eqnarray}
We may observe that $f_{C}(\rho_{AB}^{G(0)})$ varies from $[0.66667,0.99997]$ when $\lambda_{4}$ varies from $[0,1]$.\\
Thus, the assumptions on non-conditioned fidelity and conditioned fidelity are met. It can be easily verified that these assumptions still hold if we consider the state  $\rho_{AB}^{G(1)}$. This means that the GHZ state described by the density operator $\rho_{CAB}^{(1)}$ is useful for the controlled quantum teleportation.\\
Now, our task is to calculate the power of the controller when three-qubit GHZ state (\ref{ghzstate}) is shared between Alice, Bob and Charlie (controller). To estimate the power of the controller, we again consider the state $\rho_{AB}^{G(0)}$ and proceed toward the calculation of the lower bound of the power that need the following information:\\
(i) $M(\rho_{AB}^{G(0)})>1$ for $\lambda_{4}\in [0,1]$. This indicate that the state $\rho_{AB}^{G(0)}$ violate the Bell-CHSH inequality and therefore, the state is useful in conventional quantum teleportation \cite{horo4}.\\
(ii) The expectation value of the witness operator $W_{xy}^{(\phi^{+})}$ with respect to the state $\rho_{AB}^{G(0)}$ is given by
\begin{eqnarray}
Tr[W_{xy}^{(\phi^{+})}(\rho_{AB}^{G(0)})]= \frac{1}{2}-2a -\frac{\Lambda}{1.8-\lambda_{4}^{2}}
\end{eqnarray}
where $\Lambda=0.90 - 0.5\lambda_{4}^{2} + 1.2\lambda_{4}\sqrt{1-\lambda_{4}^{2}}$.\\
The value of $a~(>0)$ is chosen in such a way that the witness operator $W_{xy}^{(\phi^{+})}$ detects the state $\rho_{AB}^{G(0)}$. Thus, we find that when $a\in (0,0.232)$ \& $0.598\leq \lambda_{4} \leq 0.95 $, the witness operator detect the state $\rho_{AB}^{G(0)}$.  Therefore, the lower bound of the controller's power can be estimated as
\begin{eqnarray}
P_{CT}^{G(0)} \geq \frac{4a}{3} (1-\sqrt{M(\rho_{AB}^{G(0)})})-\frac{2}{3}Tr[W_{ij}^{(\phi^{+})}\rho_{AB}^{G(0)}] 
\end{eqnarray}
It can be easily verified that lower bound of power lies in the interval (0.001472,0.333) for $a\in (0,0.1555)$ \& $0.598\leq \lambda_{4} \leq 0.95$.\\
Also, We note that the calculation of the power of the controller for the state $\rho_{AB}^{G(1)}$ may be done in a similar way.\\
Moreover, we may consider other pure three-qubit states such as maximally slice state $|\psi^{(2)}\rangle_{ABC}= \lambda_{0}|000\rangle +\lambda_{1}|100\rangle+\frac{1}{\sqrt{2}}|111\rangle$ given in \cite{li2014} and W class states $|W_{n}\rangle=\frac{1}{\sqrt{2+2n}}(|100\rangle+\sqrt{n}|010\rangle+\sqrt{n+1}|001\rangle)$ introduced in \cite{patiagrawal}. We have analyzed the power of the controller for these classes of states, which is given in Table-I\label{t1}. We also find that the pure three-qubit W class of state described by $|W_{1}\rangle$ is more useful in CQT scheme than all other W class of states such as $|W_{2}\rangle$, $|W_{3}\rangle$ etc. $|W_{1}\rangle$ is more useful in CQT scheme in the sense that when $|W_{1}\rangle$ is used, the power of the controller is greater than all the power calculated over the states $|W_{2}\rangle$, $|W_{3}\rangle$ etc. This finding is shown in Fig.1.
\begin{figure}[h!]
	\includegraphics[width=10cm, height= 8cm]{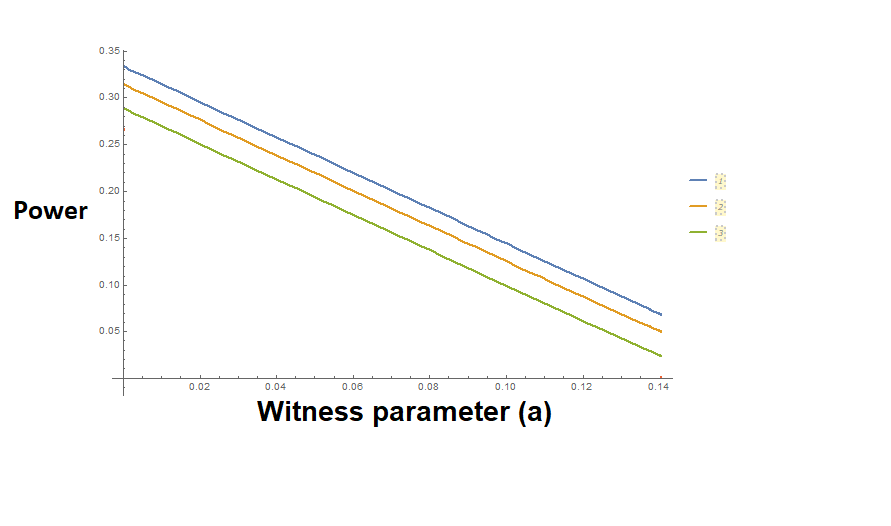}
	\caption{ This graph shows the relation between witness parameter $(a)$ and the lower limit of controller's power $P_{CT}^{W_{1}(0)}$, $P_{CT}^{W_{2}(0)}$, $P_{CT}^{W_{3}(0)}$. Blue line denotes the power of $W_{1}$ state, Yellow line denotes the power of $W_{2}$ state, and Green line denotes the power of $W_{3}$ state}
\end{figure} 
\begin{table*}[!htbp]
\begin{tabular} {|c|c|c|c|}\hline 
Three-qubit State & Non-Conditioned Fidelity & Conditioned Fidelity & Estimated lower bound of the power \\  \hline 
$|\psi^{(2)}\rangle$ & [0.5,0.6667] & \begin{tabular}{c}$f_{C}(\rho_{AB}^{MSS(0)})$=$f_{C}(\rho_{AB}^{MSS(1)})$=[0.667538,1] \\$\lambda_{4} \in [0,0.643]$  \end{tabular}  & \begin{tabular}{c}
(0,0.3333), $a \in (0,0.0402]$	\\ $\lambda_{4} \in [0,0.6]$
\end{tabular} \\ \hline 
			
\multirow{2}{*}{$|W_{1}\rangle$} & \multirow{2}{*}{$\frac{2}{3}$} & $f_{C}(\rho_{AB}^{W_{1}(0)})$= 0.9999999999968732 & [0.00146455,0.33333), a$\in$(0,0.1767]  \\ \cline{3-4}
&  & $f_{C}(\rho_{AB}^{W_{1}(1)})$=0.99999999976277 & [0.00146455,0.33333), a$\in$(0,0.1767]  \\ \hline  
$|W_{2}\rangle$ & 0.657135 &$f_{C}(\rho_{AB}^{W_{2}(0)})$= $f_{C}(\rho_{AB}^{W_{2}(1)})$=0.980937 & [0.0001815,0.31427), a$\in$(0,0.1665]  \\ \hline  	
$|W_{3}\rangle$ & 0.644337 & $f_{C}(\rho_{AB}^{W_{3}(0)})$=$f_{C}(\rho_{AB}^{W_{3}(1)})$=0.955342 & [0.000427,0.288675), a$\in$(0,0.1525]   \\ \hline		
\end{tabular}
\caption{In this table, we have estimated the lower bound of controller’s power when various three-qubit pure state such as maximally slice state $|\psi^{(2)}\rangle$, and $|W_{n}\rangle,~n=1,2,3$ states. We have found that all the three-qubit states are useful for controlled teleportation and furthermore, we obtain that $|W_{1}\rangle$ is more useful in controlled teleportation in comparison to $|W_{2}\rangle$ and $|W_{3}\rangle$ state. }
\label{t1}
\end{table*}

\section{Controlled teleportation in noisy environment}
\noindent The standard W state is given by
\begin{eqnarray}
|\psi^{(W)}\rangle_{BAC}&=& \frac{1}{\sqrt{3}}(|000\rangle +|101\rangle+ |110\rangle) 
\label{wstate}
\end{eqnarray}
It is known that the standard W state is not useful in controlled quantum teleportation \cite{Artur}. Therefore,  in this section, we investigate for the possibility of using the standard W state in CQT protocol when one of the qubit passes through the noisy environment. To execute our protocol, we assume that a source generate three-qubit entangled state $\rho^{(W)}_{BAC}=|\psi^{(W)}\rangle_{BAC}\langle \psi^{(W)}|$, where $|\psi^{(W)}\rangle_{BAC}$ is given in the form (\ref{wstate}). In this protocol, let us further assume that the two party Alice and Charlie are in one place while Bob is another distant partner. Alice possesses the two qubits $A$ and $B$ respectively. On the other hand, Charlie have the qubit $C$. Thus, initially, the qubit $B$ is also with Alice. Since Alice would like to send some information to Bob via a shared quantum state so she need to construct an entangled channel between them. Thus, Alice has to send a qubit (suppose, a qubit B) involved in the three-qubit entangled state $\rho^{(W)}_{BAC}$ to Bob through the noisy environment. The noisy environment may be described by either as (i) Amplitude Damping Channel or (ii) Phase Damping Channel. The qubit $B$ then interact with the noisy environment while travelling to Bob's place and assuming that finally it reaches to Bob. In this way, a channel is constructed between Alice and Bob through which Alice can send her information to Bob using quantum teleportation protocol. Since the qubit $B$ has interacted with the noisy environment so there may be a possibility of the degradation of the entanglement of the established channel between Alice and Bob. Thus, the teleportation fidelity may become less than $\frac{2}{3}$. In this scenario, Charlie may play a major role as controller to enhance the teleportation fidelity. Hence, we can calculate the power of the controller in this version of controlled teleportation.    
\subsection{Amplitude Damping Channel}
\noindent  Recalling the standard W state given in (\ref{wstate}) and following the above described protocol where the qubit $B$ is interacting with the noisy environment. Let us consider first the amplitude damping channel as the noisy environment through which the qubit $B$ is passing. Amplitude damping channel is described by the Kraus operators defined as \cite{ban}
\begin{eqnarray}
K_{1}=|0\rangle \langle0|+\sqrt{1-p}|1\rangle \langle1|,  
K_{2}=\sqrt{p}|0\rangle \langle1|,~~0\leq p\leq 1 \nonumber\\
\label{krauss}
\end{eqnarray} 
The Kraus operator satisfies $K_{1}^{\dagger}K_{1}+K_{2}^{\dagger}K_{2}=I$.\\
When the qubit $B$ passes through the amplitude damping channel, the state $\rho_{BAC}^{(W)}$ reduces to 
\begin{eqnarray}
\rho_{BAC}^{(W1)}&=&(K_{1}\otimes I\otimes I)\rho_{BAC}^{(W)}(K_{1}^{\dagger}\otimes I\otimes I)\nonumber \\ &+& (K_{2}\otimes I\otimes I)\rho_{BAC}^{(W)}(K_{2}^{\dagger}\otimes I\otimes I)\nonumber \\
&=& \frac{1}{3}(|000\rangle \langle000|+ p(|001\rangle \langle001|+|001\rangle \langle010|\nonumber \\ &+&|010\rangle \langle001|+|010\rangle \langle010|)+\sqrt{1-p}(|000\rangle \langle101|\nonumber \\&+&|000\rangle \langle110|+|101\rangle \langle000|+|110\rangle \langle000|) \nonumber \\&+&(1-p)(|101\rangle \langle101|+|101\rangle \langle110|+|110\rangle \langle101|\nonumber \\ &+&|110\rangle \langle110|)) 
\end{eqnarray}
Now, our task is to see whether the channel generated between Alice and Bob is useful for conventional teleportation. To verify this, we trace out the system $C$ from the state described by the density operator $\rho_{BAC}^{(W1)}$. The resulting two-qubit state is then given by
\begin{eqnarray}
\rho_{BA}^{(W1)}&=&\frac{1}{3}(|00\rangle \langle00| + p(|00\rangle \langle00|+|01\rangle \langle01|)\nonumber \\ &+&\sqrt{1-p}(|00\rangle \langle11|+|11\rangle \langle00|)\nonumber \\ &+& (1-p)(|10\rangle \langle10|+|11\rangle \langle11|))
\end{eqnarray}
The non-conditioned fidelity of teleportation when $\rho_{BA}^{(W1)}$ is used as a resource state, is given by
\begin{eqnarray}
f_{NC}(\rho_{BA}^{(W1)})&=&\frac{5+2\sqrt{1-p}}{9}\nonumber \\
&\leq& \frac{2}{3}, ~~ \text{for}~~\frac{3}{4}\leq p\leq 1
\end{eqnarray}
Therefore, we find here that there exist a range of the noisy parameter $p$ $(\frac{3}{4}\leq p \leq 1)$ for which $f_{NC}(\rho_{BA}^{(W1)})\leq \frac{2}{3}$. Thus, in this case, the controller (Charlie) has to use his power to enhance the fidelty of teleportation. To do this task, let us recall again the three-qubit state $\rho_{BAC}^{(W1)}$ ad then charlie performs von-Neumann measurement $\{B_{k},k=0,1\}$ on his qubit $C$. According to the measurement result, the resulting two-qubit states are given by  
\begin{eqnarray}
\rho_{BA}^{W1(0)}&=&\frac{1}{3N_{0}}\bigg( (t^{2}+y_{3}^{2})(|00\rangle \langle00|+p|01\rangle \langle01|\nonumber\\ &+&\sqrt{1-p}(|00\rangle \langle11|+|11\rangle \langle00|)+(1-p)|11\rangle \langle11|)\nonumber \\ &+& (-ty_{2}+y_{1}y_{3}+i(ty_{1}+y_{2}y_{3}))(p|01\rangle \langle00|\nonumber \\&+& \sqrt{1-p}|00\rangle \langle10|+ (1-p)|11\rangle \langle10| )\nonumber \\ &+& (-ty_{2}+y_{1}y_{3} -i(ty_{1}+y_{2}y_{3}))(p|00\rangle \langle01|\nonumber \\ &+& \sqrt{1-p}|10\rangle \langle00|+ (1-p)|10\rangle \langle11|)\nonumber \\ &+& (y_{1}^{2}+y_{2}^{2}) (p|00\rangle \langle00|+(1-p)|10\rangle \langle10|)\bigg) 
\end{eqnarray}
\begin{eqnarray}
	\rho_{BA}^{W1(1)}&=&\frac{1}{3N_{1}}\bigg( (y_{1}^{2}+y_{2}^{2})(|00\rangle \langle00|+p|01\rangle \langle01|\nonumber\\ &+&\sqrt{1-p}(|00\rangle \langle11|+|11\rangle \langle00|)+(1-p)|11\rangle \langle11|)\nonumber \\ &+& (ty_{2}-y_{1}y_{3}-i(ty_{1}+y_{2}y_{3}))(p|01\rangle \langle00|\nonumber \\&+& \sqrt{1-p}|00\rangle \langle10|+ (1-p)|11\rangle \langle10| )\nonumber \\ &+& (ty_{2}-y_{1}y_{3}+i(ty_{1}+y_{2}y_{3}))(p|00\rangle \langle01|\nonumber \\ &+& \sqrt{1-p}|10\rangle \langle00|+ (1-p)|10\rangle \langle11|)\nonumber \\ &+& (t^{2}+y_{3}^{2}) (p|00\rangle \langle00|+(1-p)|10\rangle \langle10|)\bigg) 
\end{eqnarray}
where $N_{0}=\frac{2(t^{2}+y_{3}^{2})+(y_{1}^{2}+y_{2}^{2})}{3}$and $N_{1}=\frac{(t^{2}+y_{3}^{2})+2(y_{1}^{2}+y_{2}^{2})}{3}$.\\
In the first case, we consider the two-qubit state $\rho_{BA}^{W1(0)}$ shared between Alice and Bob. We now choose the measurement parameter $(t,y_{1},y_{2},y_{3})$ in such a way that the fidelity of teleportation would be greater than $\frac{2}{3}$. Therefore, the measurement parameters may be chosen as
\begin{eqnarray}
&&t=0.9615239544277027\nonumber\\&& y_{1}=-0.00000006450287021375004\nonumber\\&& y_{2}=-0.000000029154369318260298\nonumber\\&& y_{3}=0.2747211110648374 
\end{eqnarray}
The conditioned fidelity is then given by
\begin{eqnarray}
f_{C}(\rho_{BA}^{W1(0)})&=& 0.666667 + 0.333333\sqrt{1-p} - 0.166667p\nonumber\\&& 0.75\leq p \leq 0.82842
\label{cfam} 
\end{eqnarray}
We may observe that the conditioned fidelity $f_{C}(\rho_{BA}^{W1(0)})$ is greater than $\frac{2}{3}$ when $0.75\leq p \leq 0.82842$. In all other range of the parameter $p$, either $f_{NC}> \frac{2}{3}$ or $f_{C}\leq \frac{2}{3}$. Thus, we will consider $0.75\leq p \leq 0.82842$ where all conditions of controlled teleportation are met. In a similar way, the condition for the controlled teleportation can be studied by considering the second case when the measurement on the charlie's qubit generate a two-qubit state described by the density operator $\rho_{AB}^{W1(1)}$. In any case, we find that both the states $\rho_{AB}^{W1(0)}$ and $\rho_{AB}^{W1(1)}$ are useful in the controlled quantum teleportation scheme. 
\subsection{Phase Damping Channel}
\noindent The phase damping channel is described by the 
Kraus Operator, which may be defined as \cite{adhikari}
\begin{eqnarray}
K_{1}&=&\sqrt{1-p}(|0\rangle \langle0|+|1\rangle \langle1|),  K_{2}=\sqrt{p}|0\rangle \langle0|,\nonumber \\
K_{3}&=&\sqrt{p}|1\rangle \langle1|,~~0\leq p\leq 1 
\end{eqnarray} 
Let us recall the standard W state described by the density operator $\rho_{BAC}^{(W)}=|\psi^{(W)}\rangle_{BAC}\langle\psi^{(W)}|$ where $|\psi^{(W)}\rangle_{BAC}$ is given in (\ref{wstate}) and follow the same protocol, as we did for amplitude damping channel. When a qubit $B$ interacted with the phase damping channel, the state $\rho_{BAC}^{(W)}$ reduces to   
\begin{eqnarray}
\rho_{BAC}^{(W2)}&=&(K_{1}\otimes I\otimes I)\rho_{BAC}^{(W)}(K_{1}^{\dagger}\otimes I\otimes I)\nonumber \\ &+& (K_{2}\otimes I\otimes I)\rho_{BAC}^{(W)}(K_{2}^{\dagger}\otimes I\otimes I)\nonumber \\  &+& (K_{3}\otimes I\otimes I)\rho_{BAC}^{(W)}(K_{3}^{\dagger}\otimes I\otimes I)\nonumber \\
&=& \frac{1}{3}\big(|000\rangle \langle000|+ |101\rangle \langle101|+|110\rangle \langle101|\nonumber \\ &+&|101\rangle \langle110|+|110\rangle \langle110|+|101\rangle \langle000|\nonumber \\ &+&|110\rangle \langle000|+|000\rangle \langle101|+|000\rangle \langle110|\nonumber \\&-& p(|101\rangle \langle000|+|110\rangle \langle000|+ |000\rangle \langle101|\nonumber \\&+&|000\rangle \langle110|)\big) 
\end{eqnarray}
To verify whether the controlled teleportation scheme is applicable for the state $\rho_{BAC}^{(W2)}$, we need to calculate non-conditioned fidelity and conditioned fidelity.\\ 
(i) Non-conditioned fidelity: The non-conditioned fidelity can be calculated as
\begin{eqnarray}
f_{NC}(\rho_{BA}^{(W2)})&=&\frac{7-2p}{9}\nonumber \\
&\leq& \frac{2}{3}, ~~ \text{for}~~\frac{1}{2}< p\leq 1
\label{ncfw2}
\end{eqnarray} 
where $\rho_{BA}^{(W2)}=Tr_{C}(\rho_{BAC}^{(W2)})$.\\
(ii) Conditioned fidelity: To calculate it, Charlie apply the measurement on his qubit in the basis $\{B_{0},B_{1}\}$. According to the measurement result, the resulting two-qubit states are given by 
\begin{eqnarray}
\rho_{BA}^{W2(0)}&=&\frac{1}{3N_{2}}\bigg( (t^{2}+y_{3}^{2})(|00\rangle \langle00|+|11\rangle \langle11|\nonumber \\ &+&|11\rangle \langle00|+|00\rangle \langle11|-p(|11\rangle \langle00|+|00\rangle \langle11|))\nonumber \\ &+&+ (-ty_{2}+y_{1}y_{3}+\iota(ty_{1}+y_{2}y_{3}))(|11\rangle \langle10|\nonumber \\ &+&|00\rangle \langle10|-p|00\rangle \langle10|)+ (-ty_{2}+y_{1}y_{3} \nonumber \\ &-&\iota(ty_{1}+y_{2}y_{3}))(|10\rangle \langle11|+|10\rangle \langle00|\nonumber \\ &-&p|10\rangle \langle00|)+ (y_{1}^{2}+y_{2}^{2}) |10\rangle \langle10|\bigg) 
\end{eqnarray}
\begin{eqnarray}
	\rho_{BA}^{W2(1)}&=&\frac{1}{3N_{3}}\bigg( (y_{1}^{2}+y_{2}^{2})(|00\rangle \langle00|+|11\rangle \langle11|\nonumber \\ &+&|11\rangle \langle00|+|00\rangle \langle11|-p(|11\rangle \langle00|+|00\rangle \langle11|))\nonumber \\ &+& (ty_{2}-y_{1}y_{3}-i(ty_{1}+y_{2}y_{3}))(|11\rangle \langle10|\nonumber \\ &+&|00\rangle \langle10|-p|00\rangle \langle10|)+ (ty_{2}-y_{1}y_{3} \nonumber \\ &+&i(ty_{1}+y_{2}y_{3}))(|10\rangle \langle11|+|10\rangle \langle00|\nonumber \\ &-&p|10\rangle \langle00|)+ (t^{2}+y_{3}^{2}) |10\rangle \langle10|\bigg) 
\end{eqnarray}
where $N_{2}=\frac{2(t^{2}+y_{3}^{2})+(y_{1}^{2}+y_{2}^{2})}{3}$and $N_{3}=\frac{(t^{2}+y_{3}^{2})+2(y_{1}^{2}+y_{2}^{2})}{3}$.\\
If the measurement parameters are given by
\begin{eqnarray}
&& t = 0.9615239543413954 \nonumber\\&&
y1 = 0.000002698965323056848 \nonumber\\&&
y2 = -0.000000004258892841348826 \nonumber\\&&
y3 = 0.2747211523762679
\end{eqnarray}
then the conditioned fidelity $f_{C}(\rho_{BA}^{W2(0)})$ is given by
\begin{eqnarray}
f_{C}(\rho_{BA}^{W2(0)})=1- 0.333333 p,~~\frac{1}{2}\leq p \leq 1
\end{eqnarray}
It may be easily verified that $f_{C}(\rho_{BA}^{W2(0)}) \in [0.66667,0.833333]$ for $p \in [0.5,1]$.
Thus, the controlled teleportation protocol may be implemented using the state $\rho_{BA}^{W2(0)}$. In the similar fashion, it may be shown that the state $\rho_{BA}^{W2(1)}$ is useful in controlled teleportation. 

\subsection{Comparision analysis of the power of the controlled teleportation}
\noindent In this section, we compare the power of the controlled teleportation when the standard W state given by (\ref{wstate}) is evolved under amplitude damping channel and phase damping channel. We will show here that the power of the conrolled teleportation in case of phase damping channel is greater than the power in case of amplitude damping channel.\\
\textbf{(a) Power of the controlled teleportation when standard W state is evolved under amplitude damping channel:} Since we find that both the state $\rho_{AB}^{W1(0)}$ and $\rho_{AB}^{W1(1)}$ are useful in the controlled teleportation scheme so we can consider any one of the state $\rho_{AB}^{W1(0)}$ or $\rho_{AB}^{W1(1)}$ to calculate the power of the controller. Let us consider the two-qubit state $\rho_{AB}^{W1(0)}$ for the estimation of the power of the controller. To estimate it, We need to calculate the following:\\
(i) The quantity $M(\rho_{AB}^{W1(0)})$ which is found out to be less than one.\\
(ii) The expectation value of the constructed witness operator $W_{ij}^{(\phi+)}$ with respect to the state $\rho_{BA}^{W1(0)}$, which is given by
\begin{eqnarray}
	Tr[W_{xy}^{(\phi^{+})}(\rho_{BA}^{W1(0)})]&=& \frac{p}{4}
	-\frac{\sqrt{1-p}}{2}+2a,\nonumber\\&&
	0.75\leq p \leq 0.82842
	\label{witam}
\end{eqnarray}
The value of $a>0$ is chosen in such a way that the witness operator $W_{xy}^{(\phi^{+})}$ detect the state $\rho_{BA}^{W1(0)}$. We find that the witness operator $W_{xy}^{(\phi^{+})}$ detects the state $\rho_{BA}^{W1(0)}$ when $a\in (0,0.005]$. \\
With all the above information, we can estimate the power ($P_{CT}^{W1(0)}$), which is given by
\begin{eqnarray}
\frac{4a}{3} (1-\sqrt{M(\rho_{BA}^{W1(0)})})-\frac{2}{3}Tr[W_{ij}^{(\phi^{+})}\rho_{BA}^{W1(0)}] \nonumber \\ \leq P_{CT}^{W1(0)} \leq \frac{1}{2}
\end{eqnarray}
We calculate the lower limit of the power $P_{CT}^{W1(0)}$ and found that the lower limit varies in the interval $[0.0056075,0.041667)$ when $a\in (0,0.005]$ \& $0.75\leq p \leq 0.8164$. \\
\textbf{(b) Power of the controlled teleportation when standard W state is evolved under phase damping channel:} In this scenario also, we find that both the state $\rho_{AB}^{W2(0)}$ and $\rho_{AB}^{W2(1)}$ are useful in the controlled teleportation scheme so we can consider any one of the state $\rho_{AB}^{W2(0)}$ or $\rho_{AB}^{W2(1)}$ to calculate the power of the controller. Let us consider the two-qubit state $\rho_{AB}^{W2(0)}$ for the estimation of the power of the controller.  The power of the controller can be estimated by 
\begin{eqnarray}
	\frac{4a}{3} (1-\sqrt{M(\rho_{BA}^{W2(0)})})-\frac{2}{3}Tr[W_{ij}^{(\phi^{+})}\rho_{BA}^{W2(0)}] \nonumber \\ \leq P_{CT}^{W2(0)} \leq \frac{1}{2} 
\end{eqnarray}
where the expectation value of the witness operatr $W_{xy}^{(\phi^{+})}$ with respect to the state $\rho_{BA}^{W2(0)}$ is given by
\begin{eqnarray}
	Tr[W_{xy}^{(\phi^{+})}\rho_{BA}^{W2(0)}]= \frac{1}{2}+2a-(1 - 0.5p) <0 ~ \text{for} \nonumber\\ a\in (0,0.035], p\in [0.5,0.859] \nonumber \\
\end{eqnarray}
Also, the quantity M($\rho_{BA}^{W2(0)}$) can be easily calculated and found out to be greater than 1. Therefore, the lower bound of power $P_{W2}^{(0)}$ lying in the interval $[0,0.16667)$ for $a\in (0,0.005]$ \& $0.5\leq p \leq 0.859$.
\begin{figure}[h!]
	\includegraphics[width=8cm, height= 8cm]{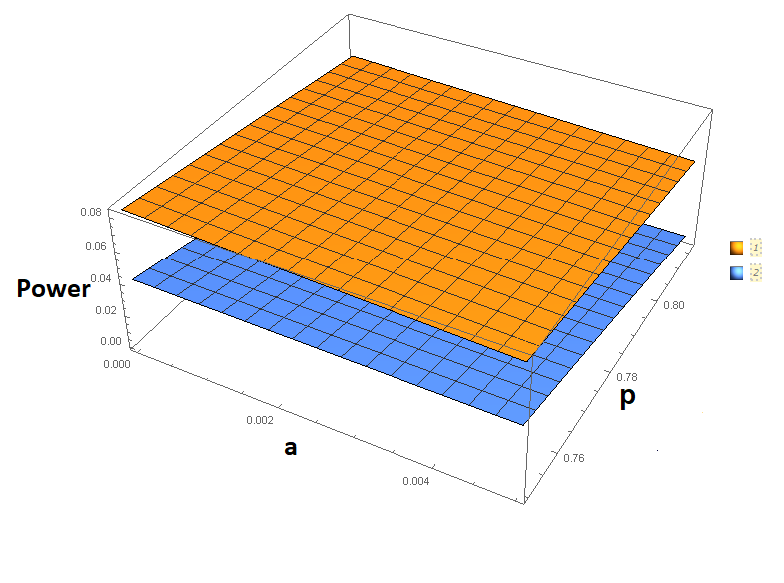}
	\caption{This graph shows the relationship between the witness parameter (a), noise parameter (p) and the controller's power. Yellow region indicates the controller's power of standard W state with phase damping channel and Blue region indicates the controller's power of standard W state with amplitude damping channel}
	\label{f2}
\end{figure} 
It can be clearly seen from Fig.-{\ref{f2}} that controller's power is more for standard $W$ state when it is under phase damping channel. Though standard $W$ state with both amplitude damping channel and phase damping channel are useful in controlled quantum channel but standard $W$ state with phase damping channel is more useful in controlled quantum teleportation. 
\section{Conclusion}
To summarize, we have considered the problem of  estimation of the power of the controller in CQT scheme. To investigate it, we have constructed a witness operator and have shown that the entangled state will be useful for teleportation as a resource state if the same entangled state is detected by the constructed witness operator and if it satisfies the Bell-CHSH inequality. Thus, at least for some cases, we need not have to use the filtering operation \cite{versatrate} to increase the teleportation fidelity. On the other hand, the study of the violation of Bell-CHSH inequality is equally important in the CQT scheme and thus we have considered the CHSH game for the estimation of the probability of success of the game through the constructed witness operator. The estimated probability of success helps in the derivation of the lower bound of the conditioned and non-conditioned fidelity in terms of the expectation value of the witness operator. Therefore, we are now able to estimate the lower and upper bound of the power of the controller in terms of the witness operator. Thus, this can pave a way to estimate the power of the controlled teleprtation in an experiment. Moreover, we have found that the state $W_{1}$ is not only useful for conventional teleportation between two parties but also useful in the CQT scheme and performs better than all the other W-class of states described by $W_{n},~n=2,3,...$ \cite{patiagrawal}. We have also studied the CQT scheme using the standard W state under a noisy environment. When one of the qubits of the standard W state passes either through the amplitude damping channel or the phase damping channel, the resulting state will be a mixed state which will be useful in controlled quantum teleportation protocol. We also found that the phase damping channel makes the controller power more positive than the amplitude damping channel. Thus, we may conclude that phase damping channel is more useful than amplitude damping channel while performing CQT protocol with the standard W state.      
\section{Acknowledgement}
\noindent A. G. would like to acknowledge the financial support from CSIR. This work is supported by CSIR File No. 08/133(0035)/2019-EMR-1.

\section{DATA AVAILABILITY STATEMENT}
Data sharing not applicable to this article as no datasets were generated or analysed during the current study.

\section{Appendix}

\subsection{Proof of Result-1}
\noindent To derive the required lower bound of the expectation value of the witness operator $W^{(\phi^{+})}_{ij}$, let us recall the witness operator defined in (\ref{witdef}). The expectation value of $W^{(\phi^{+})}_{ij}$ with respect to an entangled state $\rho_{ent}$, is given by
\begin{eqnarray}
	Tr[W^{(\phi^{+})}_{ij}\rho_{ent}]&=&(\frac{1}{2}+2a)-\langle \phi^{+}|\rho_{ent}|\phi^{+}\rangle-a\times\nonumber\\&&
	Tr[B_{ij}\rho_{ent}]\nonumber\\&\geq& (\frac{1}{2}+2a)-F(\rho_{ent})-aTr[B_{ij}\rho_{ent}]\nonumber\\&\geq&
	(\frac{1}{2}-F(\rho_{ent}))+2a(1-\sqrt{M(\rho_{ent})})\nonumber\\
	\label{lb}
\end{eqnarray}
In the second step, we have used $\langle \phi^{+}|\rho_{ent}|\phi^{+}\rangle \leq F(\rho_{ent})$. In the third step, we use the following  $Tr[B_{ij}\rho_{ent}]=\langle B_{ij}\rangle_{\rho_{ent}}\leq max_{B_{ij}}\langle B_{ij}\rangle_{\rho_{ent}}=2\sqrt{M{(\rho_{ent}})}$ for any $(i,j)$, where $i,j=x,y,z;i\neq j$ \cite{horo3}.\\ 
Let us now derive the upper bound of the expectation value of the witness operator $W^{(\phi^{+})}_{ij}$. Again, the expectation value of $Tr[W_{ij}^{(\phi^{+})}(\rho_{ent})]$ can be expressed as
\begin{eqnarray}
Tr[W_{ij}^{(\phi^{+})}\rho_{ent}] &=& \frac{1}{2}- \langle \phi^{+}|\rho_{ent}| \phi^{+}\rangle +a \times \nonumber \\&& (2-Tr[B_{ij}\rho_{ent}]) 
\label{w1}
\end{eqnarray}
Let us assume that the two qubit entangled state $\rho_{ent}$ satisfies the Bell-CHSH inequality, i.e., $Tr[B_{ij}\rho_{ent}] \in [-2,2]$ for any $i,j=x,y,z;i\neq j$. Let us split the interval $[-2,2]$ into two subintervals $[-2,0)$ and $[0,2]$. Therefore, we have the following two cases:\\
(i) If $Tr[B_{ij}\rho_{ent}] \in [0,2]$ then we get  
\begin{eqnarray}
	Tr[W_{ij}^{(\phi^{+})}\rho_{ent}] \leq \frac{1}{2}- \langle \phi^{+}|\rho_{ent}| \phi^{+}\rangle +2a  
	\label{w2}
\end{eqnarray} 
(ii) If $Tr[B_{ij}\rho_{ent}] \in [-2,0]$, we get 
\begin{eqnarray}
	Tr[W_{ij}^{(\phi^{+})}\rho_{ent}] \leq \frac{1}{2}- \langle \phi^{+}|\rho_{ent}| \phi^{+}\rangle +4a
	\label{w3}
\end{eqnarray}
Thus, combining (\ref{w2}) and (\ref{w3}) and since $a>0$, we get 
\begin{eqnarray}
	Tr[W_{ij}^{(\phi^{+})}\rho_{ent}] \leq \frac{1}{2}- \langle \phi^{+}|\rho_{ent}| \phi^{+}\rangle +4a
	\label{ub}
\end{eqnarray}
Hence, if $M(\rho_{ent})\leq 1$ then the lower and upper bound of the expectation value of the witness operator $W^{(\phi^{+})}_{ij}$ is given by
\begin{eqnarray}
&&(\frac{1}{2}-F(\rho_{ent}))+2a(1-\sqrt{M(\rho_{ent})}) \leq	Tr[W^{(\phi^{+})}_{ij}\rho_{ent}]\nonumber\\ && \leq \frac{1}{2}- \langle \phi^{+}|\rho_{ent}|\phi^{+}\rangle +4a,~~a>0
\end{eqnarray}

\subsection{Proof of Result-3}
\noindent Let us start with the lower bound of the expectation value of the witness operator $W_{ij}^{(\phi^{+})}$ $(i,j= x,y,z; i\neq j)$. Therefore, the inequality (\ref{lb}) can be re-expressed as 
\begin{eqnarray}
F(\rho_{ent}) \geq \frac{1}{2}- Tr[W_{ij}^{(\phi^{+})}\rho_{ent}] +2a(1-\sqrt{M(\rho_{ent})})\nonumber\\ 
\label{lb1}
\end{eqnarray}
The relation between the teleportation fidelity $f(\rho_{ent})$ and singlet fraction $ F(\rho_{ent})$ of an entangled state $\rho_{ent}$ is given by \cite{horo5}
\begin{eqnarray}
f(\rho_{ent})=\frac{2F(\rho_{ent})+1}{3}
\label{rel}
\end{eqnarray} 
Using (\ref{lb1}) and (\ref{rel}), we get 
\begin{eqnarray}
f(\rho_{ent}) \geq \frac{2}{3}[1- Tr[W_{ij}^{(\phi^{+})}\rho_{ent}] +2a(1-\sqrt{M(\rho_{ent})})] \nonumber\\
\label{reltel}
\end{eqnarray}
Using the fact that $M(\rho_{ent})\leq 1$ and the witness operator $W^{(\phi^{+})}_{ij}$ detect the entangled state $\rho_{ent}$, it can be easily verified that $f(\rho_{ent})>\frac{2}{3}$. Further, imposing the condition that $f(\rho_{ent}) \leq 1$, we can obtain  the upper bound of the parameter $a$, which is given by   
\begin{eqnarray}
a \leq \frac{\frac{1}{2}+Tr[W_{ij}^{(\phi^{+})}(\rho_{ent})]}{2(1-\sqrt{M(\rho_{ent})})}]
\end{eqnarray}
Therefore, the interval of the parameter $a$ for which the entangled state $\rho_{ent}$ satisfies the inequality $M(\rho_{ent})\leq 1$ and useful for teleportation is given by
\begin{eqnarray}
a \in(0,\frac{\frac{1}{2}+Tr[W_{ij}^{(\phi^{+})}(\rho_{ent})]}{2(1-\sqrt{M(\rho_{ent})})}]
\end{eqnarray}


\begin{thebibliography}{90}
\bibitem{nielsen} M. A. Nielsen, and I. L. Chuang, Quantum computation and quantum information (Cambridge University Press, Cambridge, 2000).
\bibitem{wilde} M. M. Wilde, Quantum information theory (Cambridge University Press, Cambridge, 2013).
\bibitem{bennett2} C. H. Bennett, G. Brassard, C. Crepeau, R. Jozsa, A. Peres, and
W. K. Wootters, Phys. Rev.
Lett. \textbf{70}, 1895 (1993).
\bibitem{gisin1} N. Gisin, G. Ribordy, W. Tittel, and H. Zbinden, Rev. Mod. Phys. \textbf{74}, 145 (2002).
\bibitem{briegel} H.-J. Briegel, W. Dur, J. I. Cirac, and P. Zoller, Phys. Rev. Lett. \textbf{81}, 5932 (1998).
\bibitem{gottesman} D. Gottesman, and I. L. Chuang, Nature \textbf{402}, 390 (1999).
\bibitem{bouwmeister}D. Bouwmeister, J. W. Pan, K. Mattle, M. Eible, H. Weinfurther,
and A. Zeilinger, Nature \textbf{390}, 575 (1997).
\bibitem{boschi}D. Boschi, S. Branca, F. De Martini, L. Hardy, and S. Popescu,
Phys. Rev. Lett. \textbf{80}, 1121 (1998).
\bibitem{kwiat}P. G. Kwiat, K. Mattle, H. Weinfurther, and A. Zeilinger,
Phys. Rev. Lett. \textbf{75}, 4337 (1995).
\bibitem{michler} M. Michler, K. Mattle, M. Eible, H. Weinfurther, and A.
Zeilinger, Phys. Rev. A \textbf{53}, R1209 (1996).
\bibitem{karlsson} A. Karlsson, and M. Bourennane, Phys. Rev. A \textbf{58}, 4394 (1998).
\bibitem{gao}T. Gao, F. L. Yan, and Y. C. Li, Euro. Phys. Lett. \textbf{84}, 50001 (2008).
\bibitem{li2014}X. Li, and S. Ghose, Phys. Rev. A \textbf{90}, 052305 (2014).
\bibitem{li2015} X. Li, and S. Ghose, Phys. Rev. A \textbf{91}, 012320 (2015).
\bibitem{jeong} K. Jeong, J. Kim, and S. Lee, Phys. Rev. A \textbf{93}, 032328 (2016).
\bibitem{Artur} A. Barasinski, and J. Svozilik, Phys. Rev. A \textbf{99}, 012306 (2019).
\bibitem{paulson} K. G. Paulson, and P. K. Panigrahi,  Phys. Rev. A \textbf{100}, 052325 (2019).
\bibitem{wang} T.-J. Wang, G.-Q. Yang, and C. Wang, Phys. Rev. A \textbf{101}, 012323 (2020).
\bibitem{kumar} A. Kumar, S. Haddadi, M. R. Pourkarimi,
B. K. Behera, and P. K. Panigrahi, Sci. Rep. \textbf{10}, 13608 (2020).
\bibitem{rau}R. Raussendorf, and H. J. Briegel, Phys. Rev. Lett. \textbf{86}, 5188 (2001).
\bibitem{hamdoun} H. Hamdoun, and A. Sagheer, Dig. Commun. Net. \textbf{6}, 463 (2020).
\bibitem{jun}  Z. Zhan-Jun, L. Yi-Min, and M. Zhong-Xiao, Commun. Theor. Phys. \textbf{44}, 847 (2005).
\bibitem{sango}N. Sangouard, C. Simon, H.  Riedmatten, and N. Gisin,
Rev. Mod. Phys. \textbf{83}, 33 (2011).
\bibitem{sayan} S. Gangopadhyay, T. Wang, A. Mashatan, and S. Ghose,
Phys. Rev. A \textbf{106}, 052433 (2022).
\bibitem{luo} S. Luo, Phys. Rev. A \textbf{77}, 042303 (2008).  
\bibitem{Guhne}O. Guhne, and G. Toth, Phys. Rep. \textbf{474}, 1 (2009).
\bibitem{horo3} R. Horodecki, P. Horodecki, and M. Horodecki, Phys. Lett. A \textbf{200}, 340 (1995).
\bibitem{hyllus} P. Hyllus, O. Guhne, D. Brub, and M. Lewenstein, Phys. Rev. A \textbf{72}, 012321 (2005).
\bibitem{Bennett}C. H. Bennett, D. P. Di Vincenzo, J. Smolin, and W. K. Wootters,
Phys. Rev. A \textbf{54}, 3824 (1997).
\bibitem{Mhorodecki}M. Horodecki, and P. Horodecki, Phys. Rev. A \textbf{59}, 4206 (1999).
\bibitem{horo4} R. Horodecki, M. Horodecki, and P. Horodecki, Phys. Lett. A \textbf{222}, 21 (1996). 

\bibitem{versatrate}F. Verstraete, and H. Verschelde, Phys. Rev. Lett. \textbf{90}, 097901 (2003).

\bibitem{jon} J. Oppenheim, and S. Wehner, Science \textbf{330}, 1072 (2010).
\bibitem{lee} S. Lee, J. Joo, and J. Kim, Phys. Rev. A \textbf{72}, 024302 (2005). 
\bibitem{patiagrawal}P. Agrawal, and A. Pati, Phys. Rev. A \textbf{74}, 062320 (2006).
\bibitem{ban} S. Bandyopadhyay, Phys. Rev. A \textbf{65}, 022302 (2002).
\bibitem{adhikari} S. Adhikari, I. Chakrabarty, and P. Agrawal, Quan. Inf. Comp. \textbf{12}, 0253 (2012).






\bibitem{horo5} M. Horodecki, P. Horodecki, and R. Horodecki, Phys. Rev. A \textbf{60}, 1888 (1999). 
\end{thebibliography}
\end{document}